\documentstyle[12pt]{article}

\begin{document}

\begin{titlepage}
\setcounter{page}{1}\title{Euclidean thermal spinor Green's function
in the spacetime of a straight cosmic string}
\author{B. LINET \thanks{E-mail : linet@ccr.jussieu.fr} \\
\mbox{\small Laboratoire de Gravitation et Cosmologie Relativistes} \\
\mbox{\small CNRS/URA 769, Universit\'e Pierre et Marie Curie} \\
\mbox{\small Tour 22/12, Bo\^{\i}te Courrier 142} \\
\mbox{\small 4, Place Jussieu, 75252 PARIS CEDEX 05, France}}
\maketitle
\begin{abstract}
Within the framework of the quantum field theory at finite temperature on
a conical space, we determine the Euclidean thermal spinor Green's
function for a massless spinor field. We then calculate the thermal
average of the energy-momentum tensor of a thermal bath of massless
fermions. In the high-temperature limit, we find that the straight
cosmic string does not perturb the thermal bath.

\end{abstract}
\thispagestyle{empty}
\end{titlepage}

\section{Introduction}

A straight cosmic string immersed in a thermal bath of massless bosons
must modify the thermal average of the energy-momentum tensor
\cite{smith,davies}. Within the quantum field theory at
finite temperature, we explicitly determined the Euclidean thermal
scalar Green's function on a conical space for a massless scalar field
\cite{linet1}. We then enabled us to evaluate the thermal average of
the energy-momentum tensor in the case of a conformally invariant scalar
field, in particular in the high-temperature limit.
Recently, Frolov {\em et al} \cite{frolov1} are extended this work to an
arbitrary massless scalar field and Guimar\~{a}es \cite{guimaraes1} has
furthermore assumed the existence of a magnetic flux running through
the cosmic string.

The aim of this paper is to study the gravitational influence of a
straight cosmic string on a thermal bath of massless fermions, for
instance neutrinos. To this purpose, we will determine the Euclidean
thermal spinor Green's function for a massless spin-$\frac{1}{2}$ field
within the quantum field theory at finite temperature. From this,
we will be able to calculate the thermal average of the energy-momentum
tensor of this thermal bath.

In general relativity, the spacetime describing a straight cosmic string
possesses a conical-type line
singularity \cite{vilenkin}; the metric can be written as
\begin{equation}
\label{1.1}
ds^{2}=d\rho^{2}+B^{2}\rho^{2}d\varphi^{2}+dz^{2}-dt^{2}
\end{equation}
in a coordinate system $(\rho ,\varphi ,z,t)$ with
$\rho \geq 0$ and $0\leq \varphi <2\pi$ where the constant $B$ is related
to the linear mass density $\mu$ of the cosmic string by $B=1-4G\mu$
$(0<B\leq 1)$. By performing a Wick rotation
\begin{equation}
\label{1.2}
\tau =-it
\end{equation}
metric (\ref{1.1}) takes a Riemannian form
\begin{equation}
\label{1.3}
ds^{2}=d\rho^{2}+B^{2}\rho^{2}d\varphi^{2}+dz^{2}+d\tau^{2}
\end{equation}
in which we can consider the Euclidean Green's functions.

The Euclidean spinor Green's function $S_{E\beta}$ at finite temperature
$T$ in metric (\ref{1.3}) is characterised by the property to be
antiperiodic in coordinate $\tau$ with period $\beta$ where $\beta =1/T$.
We use units in which $c=\hbar =k_{B}=1$. The spinor Green's functions
can be derived from the scalar Green's functions following a procedure
that we have already used \cite{linet2} in metric (\ref{1.3}).
In the present case, the determination of the thermal spinor Green's
function requires the knowledge of the scalar Green's function
$G_{A}^{(\gamma )}$ which is antiperiodic in coordinate $\tau$
with period $\beta$, {\em i.e.}
\begin{equation}
\label{1.4}
G_{A}^{(\gamma )}(\tau +\beta )=-G_{A}^{(\gamma )}(\tau )
\end{equation}
and which is subject to the following boundary condition
\begin{equation}
\label{1.5}
G_{A}^{(\gamma )}(\varphi +2\pi )=\exp (2\pi i \gamma )
G_{A}^{(\gamma )}(\varphi )
\end{equation}
for a constant $\gamma$ $(0\leq \gamma <1$) that we will subsequently
give the value in function of the constant $B$.

The plan of the work is as follows. In section 2, we determine the scalar
Green's function $G_{A}^{(\gamma )}$ in a handy form for a massless field,
hereafter denoted $D_{A}^{(\gamma )}$. We show in section 3 that we can
derive from this an expression of the spinor Green's function $S_{E\beta}$
which is locally the sum of the usual thermal spinor Green's function in
Euclidean space and a regular part specifically induced by the global
geometry of the spacetime of a straight cosmic string. In section 4, by
using the Wick rotation, we calculate the thermal average of the
energy-momentum of a thermal bath of massless fermions. We add some
concluding remarks in section 5.

\section{Antiperiodic scalar Green's function}

For a massive scalar field of mass $m$, the scalar Green's function
$G_{A}^{(\gamma )}$ obeys the equation
\begin{equation}
\label{2.1}
(\Box -m^{2})G_{A}^{(\gamma )}(x,x_{0};m)=-\delta^{(4)}(x,x_{0})
\end{equation}
where $\Box$ is the Laplacian operator and $\delta^{(4)}$ the Dirac
distribution in metric (\ref{1.3})
and furthermore it satisfies conditions (\ref{1.4}) and (\ref{1.5}).

Our method is based on the fact that
the scalar Green's functions can be deduced from the heat kernels
within the Schwinger-De-Witt formalism. In the present case, we have
\begin{equation}
\label{2.4}
G_{A}^{(\gamma )}(x,x_{0};m)=\int_{0}^{\infty}K_{A}^{(\gamma )}(x,x_{0};s)ds
\end{equation}
where the heat kernel $K_{A}^{(\gamma )}$ satisfies the condition
corresponding to (\ref{1.4})
\begin{equation}
\label{2.5}
K_{A}^{(\gamma )}(\tau +\beta )=-K_{A}^{(\gamma )}(\tau )
\end{equation}
and the condition corresponding to (\ref{1.5})
\begin{equation}
\label{2.6}
K_{A}^{(\gamma )}(\varphi +2\pi )=\exp (2\pi i \gamma )
K_{A}^{(\gamma )}(\varphi )
\end{equation}
We recall that the Euclidean heat kernel $K_{A}^{(\gamma )}$ obeys
the equation
\[
(\frac{\partial}{\partial s}-\Box +m^{2})K_{A}^{(\gamma )}=0 \quad (s>0)
\quad {\rm with} \quad
\lim_{s\rightarrow 0}K_{A}^{(\gamma )}(x,x_{0};s)=\delta^{(4)}(x,x_{0})
\]

For an ultrastatic metric, {\em i.e.} static and $g_{\tau \tau}=1$,
Braden \cite{braden}  has proved that $K_{A}^{(\gamma )}$ may be factorized
with the aid of the following theta function
\[
\theta_{4}(z\mid \tau )=1+\sum_{n=-\infty}^{\infty}
\exp (i\pi n^{2}\tau +2nz)
\quad (\Im \tau >0)
\]
under the form
\begin{equation}
\label{2.7}
K_{A}^{(\gamma )}(x,x_{0};m)=\theta_{4}(i\frac{\beta (\tau -\tau_{0})}{4s}
\mid i\frac{\beta^{2}}{4\pi s})K_{E}^{(\gamma )}(x,x_{0};m)
\end{equation}
where $K_{E}^{(\gamma )}$ is the zero-temperature heat kernel which vanishes
when the points $x$ and $x_{0}$  are infinitely separated. In
metric (\ref{1.3}), formula (\ref{2.4}) yields the scalar Green's function
$G_{E}^{(\gamma )}$
for a charged scalar field interacting with a magnetic flux $\gamma$;
it has been explicitly determined by Guimar\~{a}es \cite{guimaraes1}.

We now recall the expression of $K_{E}^{(\gamma )}$ because it will be
needed in the next calculations. We confine ourselves to the case $B>1/2$.
In the subset of the considered space defined by
\begin{equation}
\label{2.9}
\frac{\pi}{B}-2\pi <\varphi -\varphi_{0}<2\pi -\frac{\pi}{B}
\end{equation}
Guimar\~aes \cite{guimaraes1} has found an expression of
$K_{E}^{(\gamma )}$ which is the sum of the usual heat kernel
\begin{equation}
\label{2.10}
K_{E}^{{\rm usuel}}(x,x_{0};s)=\frac{1}{16\pi^{2}s^{2}}
\exp (-\frac{r_{4}^{2}}{4s}-m^{2}s)
\end{equation}
with $r_{4}=\sqrt{(\tau -\tau_{0})^{2}+(z-z_{0})^{2}+\rho^{2}+rho_{0}^{2}
-2\rho \rho_{0}\cos [B(\varphi -\varphi_{0})]}$ and a regular part which
has the integral expression
\begin{equation}
\label{2.11}
K_{E}^{(\gamma )*}(x,x_{0};s)=\frac{\exp (-m^{2}s)}{32\pi^{3}Bs^{2}}
\int_{0}^{\infty}\exp [-\frac{R_{4}^{2}(u)}{4s}]
F_{B}^{(\gamma )}(u,\varphi -\varphi_{0})du
\end{equation}
with $R_{4}(u)=\sqrt{(\tau -\tau_{0})^{2}+(z-z_{0})^{2}+\rho^{2}
+\rho_{0}^{2}+2\rho \rho_{0}\cosh u}$ and where the function
$F_{B}^{(\gamma )}(u,\psi )$
has been determined \cite{guimaraes1,guimaraes2} under the form
\begin{eqnarray}
\label{2.11a}
\nonumber & &F_{B}^{(\gamma )}(u,\psi )=i\frac{e^{i(\psi +\pi /B)\gamma}
\cosh [u(1-\gamma )/B]-e^{-i(\psi +\pi /B)(1-\gamma )}\cosh [u\gamma /B]}
{\cosh (u/B)-\cos (\psi +\pi /B)} \\
& &-i\frac{e^{i(\psi -\pi /B)\gamma}\cosh [u(1-\gamma )/B]
-e^{-i(\psi -\pi /B)(1-\gamma )}\cosh [u\gamma /B]}
{\cosh (u/B)-\cos (\psi -\pi /B)}
\end{eqnarray}

We can now obtain the desired expression of $G_{A}^{(\gamma )}$ valid in
subset (\ref{2.9}). In the case $m=0$, it has a handy form. By substituting
(\ref{2.7}) into (\ref{2.4}), we obtain $D_{A}^{(\gamma )}$ as the sum of
the antiperiodic scalar Green's function in an Euclidean space
\begin{equation}
\label{2.12}
D_{A}^{{\rm usuel}}(x,x_{0})=\int_{0}^{\infty}
\theta_{4}(i\frac{\beta (\tau -\tau_{0})}{4s}
\mid i\frac{\beta^{2}}{4\pi s})\frac{1}{4s^{2}}
\exp (-\frac{r_{4}^{2}}{4s})ds
\end{equation}
and a regular part which has the integral expression
\begin{eqnarray}
\label{2.13}
\nonumber & &D_{A}^{(\gamma )*}(x,x_{0})=\frac{1}{32\pi^{3}B}
\int_{0}^{\infty}\int_{0}^{\infty}
\theta_{4}(i\frac{\beta (\tau -\tau_{0})}{4s}
\mid i\frac{\beta^{2}}{4\pi s}) \\
& &\times \frac{1}{s^{2}}\exp [-\frac{R_{4}^{2}(u)}{4s}]
F_{B}^{(\gamma )}(u,\varphi -\varphi_{0})duds
\end{eqnarray}
By changing the variable of integration $x=1/s$, we rewrite (\ref{2.13}) as
\begin{eqnarray}
\label{2.14}
\nonumber & &D_{A}^{(\gamma )*}(x,x_{0})=\frac{1}{32\pi^{3}B}
\int_{0}^{\infty}\int_{0}^{\infty}
\theta_{4}(i\frac{\beta (\tau -\tau_{0})}{4}x
\mid i\frac{\beta^{2}}{4\pi}x) \\
& &\times \exp [-\frac{R_{4}^{2}(u)}{4}x]
F_{B}^{(\gamma )}(u, \varphi -\varphi_{0})dxds
\end{eqnarray}
The $x$-integration in expression (\ref{2.14}) can be performed by using
the formula
\[
\int_{0}^{\infty}\theta_{4}(i\pi kx\mid i\pi x)\exp [-(k^{2}+l^{2})x]dx
=\frac{\sinh l}{l(\cosh l-\cos k)}-\frac{\sinh 2l}{l(\cosh 2l-\cos 2k)}
\]
This result is proved with the help of the identity between the
theta functions
\[
\theta_{4}(z\mid \tau )=2\theta_{3}(2z\mid 4\tau )-\theta_{3}(z\mid \tau )
\]
already noticed by Dowker and Schofield \cite{dowker},
and then by using the following integral
\[
\int_{0}^{\infty}\theta_{3}(i\pi kx\mid i\pi x)\exp [-(k^{2}+l^{2})x]dx
=\frac{\sinh 2l}{l(\cosh 2l-\cos 2k)}
\]
that we have already considered \cite{linet1}.
Thus, the expression of the antiperiodic scalar Green's function
$D_{A}^{(\gamma )}$, valid in subset (\ref{2.9}), is the sum of the
antiperiodic scalar Green's function in an Euclidean space
\begin{eqnarray}
\label{2.15}
\nonumber & &D_{A}^{{\rm usuel}}(x,x_{0})=\frac{1}{4\pi \beta r_{3}}
\{ \frac{\sinh (\pi r_{3}/\beta )}{\cosh (\pi r_{3}/\beta)
-\cos [\pi (\tau -\tau_{0})/\beta ]} \\
& &-\frac{\sinh (2\pi r_{3}/\beta )}{\cosh (2\pi r_{3}/\beta )
-\cos [2\pi (\tau -\tau_{0})/\beta ]}\}
\end{eqnarray}
with $r_{3}=\sqrt{(z-z_{0})^{2}+\rho^{2}+\rho_{0}^{2}-2\rho \rho_{0}\cos
[B(\varphi -\varphi_{0})]}$ and a regular part which has the
integral expression
\begin{eqnarray}
\label{2.16}
\nonumber & &D_{A}^{(\gamma )*}(x,x_{0})=\frac{1}{8\pi^{2}\beta B}
\int_{0}^{\infty}\frac{1}{R_{3}(u)}\{
\frac{\sinh [\pi R_{3}(u)/\beta]}{\cosh [\pi R_{3}(u)/\beta ]-\cos [\pi
(\tau -\tau_{0})/\beta ]} \\
& &-\frac{\sinh [2\pi R_{3}(u)/\beta]}{\cosh [2\pi R_{3}(u)/\beta ]
-\cos [2\pi (\tau -\tau_{0})/\beta]}
F_{B}^{(\gamma )}(u,\varphi -\varphi_{0})du
\end{eqnarray}
with
$R_{3}(u)=\sqrt{(z-z_{0})^{2}+\rho^{2}+\rho_{0}^{2}+2\rho \rho \cosh u}$
where $F_{B}^{(\gamma )}(u,\psi )$ is given by (\ref{2.11a}).

\section{Thermal spinor Green's function}

In order to write down the Dirac operator in metric (\ref{1.3}),
we choose the following vierbein
\begin{equation}
\label{3.1}
e_{\underline{2}}^{\mu}=(0,\frac{1}{B\rho},0,0) \quad
e_{\underline{a}}^{\mu}=\delta_{\underline{a}}^{\mu} \quad
\underline{a}\neq 2
\end{equation}
which is different of one often used \cite{jackiw}.
In the study of the vacuum polarization, we determined the ordinary spinor
Green's function $S_{E}$ with this choice \cite{linet2}. The thermal spinor
Green's function $S_{E\beta}$ for a massive spin-$\frac{1}{2}$ field obeys
the equation
\begin{equation}
\label{3.2}
(e_{\underline{a}}^{\mu}\gamma^{\underline{a}}
+\frac{\gamma^{\underline{1}}}
{2\rho}+mI)S_{E\beta}(x,x_{0};m)=-I\delta^{(4)}(x,x_{0})
\end{equation}
where the $\gamma^{\underline{a}}$ are the Dirac matrices such that
$\gamma^{\underline{a}\dag}=-\gamma^{\underline{a}}$ and $I$ is the
unit matrix.  As explained in \cite{linet2}, due to choice (\ref{3.1})
of vierbein, the spinor Green's function is well defined if we impose
the boundary condition
\begin{equation}
\label{3.3}
S_{E\beta}(\varphi +2\pi )=-S_{E\beta}(\varphi )
\end{equation}
The thermal character is expressed by the property of antiperiodicity in
the coordinate $\tau$ with period $\beta$, {\em i.e.}
\begin{equation}
\label{3.4}
S_{E\beta}(\tau +\beta )=-S_{E\beta}(\tau )
\end{equation}

The spinor Green's function $S_{E\beta}$ satisfying
(\ref{3.2}), (\ref{3.3}) and (\ref{3.4}) can be written as
\begin{equation}
\label{3.5}
S_{E\beta}(x,x_{0};m)=(e_{\underline{a}}^{\mu}\gamma^{\underline{a}}
\partial_{\mu}
+\frac{\gamma^{\underline{1}}}{2\rho}-mI){\cal G}_{A}(x,x_{0};m)
\end{equation}
where we have set
\begin{equation}
\label{3.6}
{\cal G}_{A}(x,x_{0};m)=(I\Re
+\gamma^{\underline{1}}\gamma^{\underline{2}}
\Im )[\exp (iB\frac{\varphi -\varphi_{0}}{2})G_{A}^{(\gamma )}(x,x_{0};m)]
\end{equation}
in which $G_{A}^{(\gamma )}$ is the antiperiodic scalar Green's function
for a parameter $\gamma$ given by the equation
\begin{equation}
\label{3.7}
\gamma =\frac{1}{2}-\frac{B}{2}
\end{equation}
We have $0\leq \mu <1/4$ since $1/2<B\leq 1$. Hence, we will obtain
in subset (\ref{2.9}) the thermal spinor Green's function under the form
\begin{equation}
\label{3.8}
S_{E\beta}(x,x_{0};m)=S_{E\beta}^{{\rm usuel}}(x,x_{0};m)
+S_{E\beta}^{*}(x,x_{0};m)
\end{equation}
where $S_{E\beta}^{*}$ can be calculated from formulas (\ref{3.5}) and
(\ref{3.6}) when one knows $G_{A}^{(\gamma )*}$ or
$D_{A}^{(\gamma )*}$ in the case $m=0$. This regular part is specifically
induced by the global geometry of the space describing a conical-type
line singularity.
In the limit where $\beta$ tends to the infinity, {\em i.e.} at
zero temperature, $S_{E\infty}$ coincides with $S_{E}$.

\section{Thermal average of the energy-momentum tensor}

Within the quantum field theory at finite temperature, a renormalization
is performed by removing the zero-temperature Green's function.
The thermal average of the energy-momentum tensor of a thermal bath of
massless fermions can be thus computed by the formula
\begin{equation}
\label{4.0}
<T_{\mu \nu}(x)>_{\beta}={\cal T}_{\mu \nu}^{(1/2)}(S_{E\beta}(x,x_{0}
-S_{E\infty}(x,x_{0})) \mid_{x=x_{0}}
\end{equation}
where ${\cal T}_{\mu \nu}^{(1/2)}$ is the following differential operator
in $x$ and $x_{0}$
suivant
\begin{equation}
\label{4.1}
{\cal T}_{\mu \nu}^{(1/2)}=\frac{1}{4}tr[\gamma^{\underline{a}}(
e_{\underline{a}\mu}(\partial_{\nu}-\partial_{\nu_{0}})+
e_{\underline{a}\nu}(\partial_{\mu}-
\partial_{\mu_{0}}))]
\end{equation}
in which we need to know only the Green's function for points
$x$ and $x_{0}$ sufficiently near.
By considering form (\ref{3.8}) of the spinor Green's function, the
energy-momentum tensor can be written down as the sum of the usual
form in an Euclidean space
\begin{equation}
\label{4.1a}
<T_{\mu \nu}(x)>_{\beta}^{{\rm usuel}}={\cal T}_{\mu \nu}^{(1/2)}
(S_{E\beta}^{{\rm usuel}}(x,x_{0})-S_{E\infty}(x,x_{0}))\mid_{x=x_{0}}
\end{equation}
and a regular term
\begin{equation}
\label{4.2}
<T_{\mu \nu}(x)>_{\beta}^{*}={\cal T}_{\mu \nu}^{(1/2)}S_{E\beta}^{*}
(x,x_{0})\mid_{x=x_{0}}
\end{equation}
which results from the existence of the conical defect.

Since the energy-momentum tensor of a thermal bath for massless
spin-$\frac{1}{2}$ field is traceless and conserved
in metric (\ref{1.1}), we get thereby
\begin{equation}
<T_{\mu}^{\nu}(x)>_{\beta}=<T_{tt}(x)>_{\beta}{\rm diag}(-1,\frac{1}{3},
\frac{1}{3},\frac{1}{3})
\end{equation}
where the energy density $<T_{tt}>$ is given by $-<T_{\tau \tau}>$
by virtue of the Wick rotation (\ref{1.2}).
By substituting (\ref{3.5}) into (\ref{4.0}), we obtain  the expression
of the energy density
\begin{equation}
\label{4.3}
<T_{tt}(x)>_{\beta}=-4\partial_{\tau \tau}\Re D_{A}^{(\gamma )}(x,x_{0})
\mid_{x=x_{0}}
\end{equation}

The application of formula (\ref{4.1a}) gives the usual energy-momentum
tensor for a thermal bath of massless fermions in an Euclidean space.
By observing that
\[
D_{A}^{{\rm usuel}}(\tau -\tau_{0})\mid_{\rho =\rho_{0},\varphi
=\varphi_{0},z=z_{0}} \sim
\frac{1}{4\pi^{2}(\tau -\tau_{0})^{2}}-\frac{1}{24\beta^{2}}
-\frac{7\pi^{2}(\tau -\tau_{0})^{2}}{480\beta^{4}}
\]
when $(\tau -\tau_{0})\rightarrow 0$, we find from (\ref{4.3}) that
the energy density is
\begin{equation}
\label{4.3a}
<T_{tt}(x)>_{\beta}^{{\rm usuel}}=\frac{7\pi^{2}}{60\beta^{4}}
\end{equation}

The application of formula (\ref{4.2}) gives the perturbation of the
energy-momentum tensor due to the straight cosmic string. From (\ref{4.3}),
we find
\begin{eqnarray}
\label{4.4}
\nonumber & &<T_{tt}(x)>_{\beta}^{*}=\frac{1}{4\beta^{3}\rho B}
\int_{0}^{\infty}\frac{1}{\cosh (u/2)} \\
& &\times \{ \frac{\sinh \frac{2\pi \rho \cosh (u/2)}{\beta}}
{[\cosh \frac{2\pi \rho \cosh (u/2)}{\beta}-1]^{2}}
-4\frac{\sinh \frac{4\pi \rho \cosh (u/2)}{\beta}}
{[\cosh \frac{4\pi \rho \cosh (u/2)}{\beta}-1]^{2}} \}
F_{B}^{(\gamma )}(u,0)du
\end{eqnarray}
in which $F_{B}^{(\gamma )}(u,0)$ is given by (\ref{2.11a}) with $\psi =0$
We recall that we have assumed that $1/2<B\leq 1$.

We firstly examine the limit of result (\ref{4.4}) when
$\beta \rightarrow \infty$, {\em i.e.} at zero temperature. By setting
\[
x=\frac{2\pi \rho \cosh (u/2)}{\beta}
\]
it becomes
\begin{eqnarray}
\nonumber & &<T_{tt}(x)>_{\beta}^{*}=\frac{1}{32\pi^{3}\rho^{4}B} \\
\nonumber & &\times \int_{0}^{\infty}
\frac{1}{[\cosh (u/2)]^{4}}x^{3}[\frac{\sinh x}{(\cosh x-1)^{2}}
-4\frac{\sinh 2x}{\cosh 2x-1)^{2}}]F_{B}^{(\gamma )}(u,0)du
\end{eqnarray}
When $\beta \rightarrow \infty$, we easily see that
\begin{equation}
\label{4.5}
<T_{tt}(x)>_{\infty}^{*}=\frac{1}{16\pi^{3}\rho^{4}}\int_{0}^{\infty}
\frac{1}{[\cosh (u/2)]^{4}}F_{B}^{(\gamma )}(u,0)du
\end{equation}
But contribution (\ref{4.3a}) to the energy density of the thermal bath
disappears at zero temperature, therefore part (\ref{4.5})  gives the
vacuum energy-momentum tensor within the vacuum polarization
which agrees with the one of Frolov and
Serebriany \cite{frolov2} as one can prove it \cite{dowker2,guimaraes2}.

We secondly examine the asymptotic behavior of result (\ref{4.4}) when
$\beta \rightarrow 0$, {\em i.e.} in the high-temperature limit. Taking
into account the properties of the hyperbolic functions, we see that
\begin{eqnarray}
\nonumber & &<T_{tt}(x)>_{\beta}^{*}\sim \frac{1}{4\beta^{3}\rho B}
\int_{0}^{\infty}  \frac{1}{\cosh (u/2)} \\
\nonumber & & \times \{ \exp [-\frac{2\pi \rho \cosh (u/2)}{\beta}]
-4\exp [-\frac{4\pi \rho \cosh (u/2)}{\beta}]\} F_{B}^{(\gamma )}(u,0)du
\end{eqnarray}
when $\beta \rightarrow 0$. Thus, the energy density
$<T_{tt}>_{\beta}^{*}$ is exponentially
decreasing in the variable $1/\beta$;  only contribution (\ref{4.3a}) to
energy density
of the thermal bath of massless fermions survives.

\section{Conclusion}

In the spacetime of a straight cosmic string, we have in fact derived a
general expression of the Euclidean thermal Green's function for a
massive spin-$\frac{1}{2}$ field but we have really got a handy form
for a massless field. We have calculated the thermal average of
the energy-momentum tensor of a thermal bath
in the case of a massless spinor field.

There exists a sharp contrast between the case of thermal bath of massless
bosons and the present situation: the energy-momentum tensor of the
thermal bath of massless fermions is not perturbed in high-temperature
limit by the presence of the straight cosmic string.

\end{document}